\documentclass[submission, Proceedings]{scipost}

\begin{document}
	
	
	\begin{center}{\Large \textbf{
				Open strange meson $K^\pm_1$ in hot and dense  nuclear matter
	}}\end{center}
	
	\begin{center}
		Rajesh Kumar\textsuperscript{1*},
		Arvind Kumar\textsuperscript{1},
	\end{center}
	
	\begin{center}
		{\bf 1} Dr B. R. Ambedkar National Institute of Technology Jalandhar, India-144011
		\\
		* rajesh.sism@gmail.com
	\end{center}
	
	\begin{center}
		\today
	\end{center}
	
	
	\section*{Abstract}
	{\bf
		Using the unification of the chiral SU(3) model and QCD sum rules,  we deduce the in-medium properties of $K^\pm_1$ meson. Within the chiral SU(3) model,
		medium modified gluon and quark condensates are evaluated through their interactions with the scalar fields ($\sigma$, $\zeta$, $\delta$, and $\chi$). These condensates are further used as input in the Borel transformed equations of QCD sum rules to evaluate the in-medium mass of strange $K^\pm_1$ meson. The in-medium property of the above meson can be used to study the restoration of chiral symmetry in nuclear matter.
	}

	\vspace{10pt}
	\noindent\rule{\textwidth}{1pt}
	\tableofcontents\thispagestyle{fancy}
	\noindent\rule{\textwidth}{1pt}
	\vspace{10pt}

	\section{Introduction}
	\label{sec:intro}
	
	The quark condensates act as an order parameter of the QCD  chiral symmetry  \cite{Song2019}. In non-perturbative QCD, the non-zero condensates lead to the spontaneously broken chiral symmetry which  also explains the mass difference between the vector  and  the partner axial-vector meson. This implies that the mass of these mesons is the same in the chiral restored phase \cite{Song2019}.  The axial-vector $K_1$(1270) meson is the partner of vector $K^*$(892) meson and in this article, we are focusing on the medium modified mass-shift of $K^+_1(u \bar s)$ and $K^+_1( \bar u s)$ mesons in the hot symmetric nuclear medium. To compute the mass-shift, we employ the union of QCD sum rules \cite{Song2019}  and chiral hadronic model \cite{Kumar2020a}. The effective hadronic mean-field model incorporates  the fundamental QCD features, for example   non-linear realization of the chiral symmetry and trace anomaly \cite{Kumar2020a}. It  has been  used in  literature to  investigate the medium induced mass splitting across charged partners of pseudoscalar $K$ (self-consistent chiral SU(3) model) \cite{Kumar2020b} and $D$ meson (chiral SU(3) model + QCD sum rules) \cite{Kumar2020a}. In this paper,  the medium (density and temperature) dependent quark (up and strange) and gluon condensates  are computed from the chiral hadronic model, which is used further in QCD sum rules  to calculate the in-medium mass of $K_1$ mesons.  Within the chiral hadronic model, the above mentioned condensates are evaluated using the in-medium  scalar  fields \cite{Chhabra2017}.  In the next segment, we show the brief methodology used in the current article.

	\section{Formalism}
	\label{sec:2}
	
	Using operator product expansion (OPE), the current-current correlator up to dimension 6 can be written as \cite{Song2019}
	\begin{align}
		\Pi(q^2)=B_0 Q^2\ln \frac{Q^2}{\mu^{2} }+ B_2\ln \frac{Q^2}{\mu^2}
		-\frac{B_4}{Q^2}-\frac{B_6}{Q^4},
		\label{OPE}
	\end{align}
	with $Q^2 \equiv -q^2$, $\mu$=1 GeV as renormalization scale and $B_n$ as Borel coefficients \cite{Song2019}. In the nuclear medium, the degeneracy of $K^+_1$ and $K^-_1$ do not hold, consequently, the charge symmetry breaking leads to even and odd contributions from the correlator, given as
	\begin{align}
		\Pi(q^2)=\Pi^e(q^2)+q_0 \Pi^o(q^2).\label{pi1eo-m}
	\end{align}

	 For small nuclear density ($\rho_N$), i.e., keeping only the linear terms in $\rho_N$, using Borel transformation the correlator can be expressed as \cite{Song2019}
	\begin{align}
		O(M^2)O'_\pm +M(M^2)m'_\pm +S(M^2){s'}_0^\pm =C_\pm(M^2),
		\label{eq:SR-m}
	\end{align}
	
	with
		\begin{align}
		&O(M^2)= -m_{K_1}^2e^{-m_{K_1}^2/M^2}, ~~ M(M^2)= O_{K_1}m_{K_1}\bigg(-\frac{3}{2}+\frac{2m_{K_1}^2}{M^2}\bigg)e^{-m_{K_1}^2/M^2},\nonumber\\
		&S(M^2)= \frac{1}{2}\bigg(1+\frac{m_{K_1}}{\sqrt{s_0}}\bigg)(B_0 s_0-B_2)e^{-s_0/M^2},~~C_\pm(M^2)=-m_s \langle\bar{u}u\rangle_N
		+\frac{\alpha_s}{12\pi}\langle G^2\rangle_N\nonumber\\
		&+\frac{m_N}{2}(A_2^u +A_2^s)\pm \frac{m_{K_1}}{3}(A_1^u-A_1^s)+\frac{32\pi \alpha_s}{9 M^2}\bigg\{
		\langle\bar{u}u\rangle_N \langle\bar{s}s\rangle_0 +\langle\bar{u}u\rangle_0 \langle\bar{s}s\rangle_N+\frac{2}{9}(\langle\bar{u}u\rangle_N \langle\bar{u}u\rangle_0\nonumber \\&+\langle\bar{s}s\rangle_N\langle\bar{s}s\rangle_0)\bigg\}-\frac{5m_N^3}{6 M^2}(A_4^u +A_4^s)\mp \frac{2m_{K_1} m_N^2}{3M^2}(A_3^u-A_3^s)+m'_\mp \bigg\{\frac{O_{K_1}m_{K_1}}{2}e^{-m_{K_1}^2/M^2}\bigg\}\nonumber\\
		&~~~+{s'}_0^\mp \bigg\{\frac{1}{2}\bigg(-1+\frac{m_{K_1}}{\sqrt{s_0}}\bigg)(B_0 s_0-B_2)e^{-s_0/M^2}\bigg\}.
		\label{minimum1}
	\end{align}

	The  in-medium mass ($m^*_\pm$), overlapping strength ($O^*_\pm$)  and  threshold parameter ($s_0^\pm{^*}$)  of $K_1^\pm$ mesons can be expressed via relations \cite{Song2019}
	

	\begin{align}
	O^*_\pm &= O_{K_1}+\Delta O^*_{K_1^\pm}= O_{K_1}+O'_\pm \rho_N, \nonumber\\
	m^*_\pm &= O_{K_1}+\Delta m^*_{K_1^\pm} =m_{K_1} +m'_\pm  \rho_N, \nonumber\\
	s_0^{\pm^*} &= s_0+\Delta  {s^*_0}^\pm= s_0+{s'}_0^\pm \rho_N,
	\label{mfms}
\end{align}

where $O_{K_1}$, $m_{K_1}$ and $s_0$ denote the vacuum values of overlapping strength, mass and threshold parameter for $K_1$ meson. In Eq. \ref{minimum1}, $A^u_i$ parameters are calculated using the MSTW parton distribution function at $\mu^2$=1 GeV$^2$ \cite{Song2019}. Also,  in the same equation, the in-medium nucleon expectation values of up  $\langle \bar u u \rangle_N$, strange quark $\langle \bar s s \rangle_N$ and gluon condensates $\langle G^2 \rangle_N$ are computed using the definition $	\langle{\cal{O}}\rangle_{N} = \frac{2m_N}{\rho_N}\left( \langle{\cal{O}}\rangle_{\rho_N}-\langle {\cal{O}}\rangle_{vac}\right)$ through the  density dependent condensates from the chiral SU(3) model \cite{Chhabra2017}. The  quark condensates  can be associated to explicit symmetry breaking  \cite{Kumar2020a}
	\begin{equation}
		\sum_{i} m_{i}\langle \bar{q}_{i}q_{i}\rangle_{\rho_N}=-\mathcal{L}_{\text{ESB}},
		\label{eqsb}
	\end{equation}
	with
	\begin{equation}
		{\cal L} _{\text{ESB}}  =  -\frac{\chi^2}{\chi_0^2}  \left [ \frac{1}{2} m_{\pi}^{2} f_{\pi} (\sigma+\delta) +\frac{1}{2} m_{\pi}^{2} f_{\pi} (\sigma-\delta) 
	+\Big( \sqrt{2} m_{K}^{2}f_{K} - \frac{1}{\sqrt{2}} m_{\pi}^{2} f_{\pi} \Big) \zeta \right ]. 
	\end{equation}
  Using Eq. \ref{eqsb},  the light quark condensates can be expressed as

	\begin{align}
		\left\langle \bar{u}u\right\rangle_{\rho_N}
		= \frac{1}{m_{u}}\left( \frac {\chi}{\chi_{0}}\right)^{2}
		\left[ \frac{1}{2} m_{\pi}^{2}
		f_{\pi} \left( \sigma + \delta \right) \right],
		\left\langle\bar{s} s\right\rangle_{\rho_N}=\frac{\chi^{2}}{\chi_{0}^{2} m_{s}}\left(\sqrt{2} m_{K}^{2} f_{K}-m_{\pi}^{2} f_{\pi}\right) \zeta.
		\label{eq:qs}
	\end{align}
	
	In above $m_u$  and $m_s$ denote the mass of up, and strange quark. Furthermore, the scalar gluon condensate $\left\langle G^2
\right\rangle_{\rho_N}$,  given by 
\begin{align}
	\left\langle G^2
	\right\rangle_{\rho_N}= \frac{8}{9} \Bigg [(1 - d) \chi^{4}
+\left( \frac {\chi}{\chi_{0}}\right)^{2} 
\left( m_{\pi}^{2} f_{\pi} \sigma
+ \big( \sqrt {2} m_{k}^{2}f_{k} - \frac {1}{\sqrt {2}} 
m_{\pi}^{2} f_{\pi} \big) \zeta \right) \Bigg ],
\nonumber \\
\label{eq:glum}
\end{align}

 is extracted using scalar fields of chiral hadronic model \cite{Kumar2020a}. Further, from Eq. (\ref{eq:SR-m}), we define the function $F_\pm(O'_\pm, m'_\pm, {s'}_0^\pm)$  as
	\begin{align}
		&F_\pm(O'_\pm, m'_\pm, {s'}_0^\pm)\equiv
		\int_{M^2_{\rm min}}^{M^2_{\rm max}} \bigg\{O(M^2)O'_\pm +M(M^2)m'_\pm+S(M^2){s'}_0^\pm - C_\pm(M^2)\bigg\}^2 dM^2,
		\label{minimum2}
	\end{align}
	with $M^2_{min}$ and $M^2_{max}$, as the lower and upper limit of the Borel window \cite{Song2019}. The following six simultaneous  linear equation for  $O'_\pm, m'_\pm, {s'}_0^\pm$ are obtained

	\begin{align}
		&O'_\pm \int dM^2 F^2(M^2) +m'_\pm \int dM^2 O(M^2)M(M^2)
		+{s'}_0^\pm\int dM^2 O(M^2)S(M^2) \nonumber\\ &= \int dM^2 O(M^2) C_\pm(M^2), \nonumber\\
		&O'_\pm \int dM^2 O(M^2)M(M^2) +m'_\pm \int dM^2 M^2(M^2)
		+{s'}_0^\pm \int dM^2 M(M^2)S(M^2)\nonumber\\ &= \int dM^2 M(M^2) C_\pm(M^2), \nonumber\\
		&O'_\pm \int dM^2 O(M^2)S(M^2) +m'_\pm \int dM^2 M(M^2)S(M^2)
		+{s'}_0^\pm \int dM^2 S^2(M^2)\nonumber\\ &= \int dM^2 S(M^2) C_\pm(M^2),
		\label{coupled1}
	\end{align}
	
	using the minimization of  function $F_\pm$, i.e., 
	
	\begin{align}
		\frac{\partial F_\pm}{\partial O'_\pm }=\frac{\partial F_\pm}{\partial m'_\pm }=\frac{\partial F_\pm}{\partial {s'}_0^\pm}=0.
	\end{align}
	The in-medium mass of $K_1^+$ and $K_1^-$ along with  $O'_\pm$ and $s'^\pm_0$ are solved simultaneously using Eq. (\ref{coupled1}) under the influence of density-dependent quark and gluon condensates.

	\section{Numerical Results and Discussions}
	\label{sec:rd}

	In this segment,  the  in-medium mass difference across the charged $K_1^+ (u \bar s)$ and $K_1^- (\bar u s)$ is discussed. In the present work, the medium modified $K_1$ mass is computed from the Borel transformed current correlator in the QCD sum rules at zero width approximation \cite{Song2019}. The density dependent  up quark, strange quark and gluon condensates are calculated in chiral  hadronic model \cite{Chhabra2017}, and are plugged in the calculations of QCD sum rules \cite{Song2019}. In the calculations of Borel sum rules, the minimum $M^2_{min}$ and maximum $M^2_{max}$ limit of Borel window is taken to be 1.06 and 2.17 GeV$^2$, respectively \cite{Song2019}. Also, the different parameters utilized in the current investigation are tabulated in table \ref{table1}.
	
	\begin{table}
		\centering
		\begin{tabular}{ccccccc}
			\hline
			$m_{K_1}$ (GeV)	& $O_{K_1}$ (GeV$^2$)  & $s_0$ (GeV$^2$)& $m_u$ (GeV)& $m_s$ (GeV)& $m_N$ (GeV) &$\alpha_s$  \\
			\hline
			1.270& 0.048 & 2.4 &0.0426 &0.117&0.940&0.5\\
			
			\hline
		\end{tabular}
		\caption{Various parameters used in the present work \cite{Song2019}.}
		\label{table1}
	\end{table}
	
		\begin{figure}
		\centering
		\includegraphics[width=0.7\linewidth]{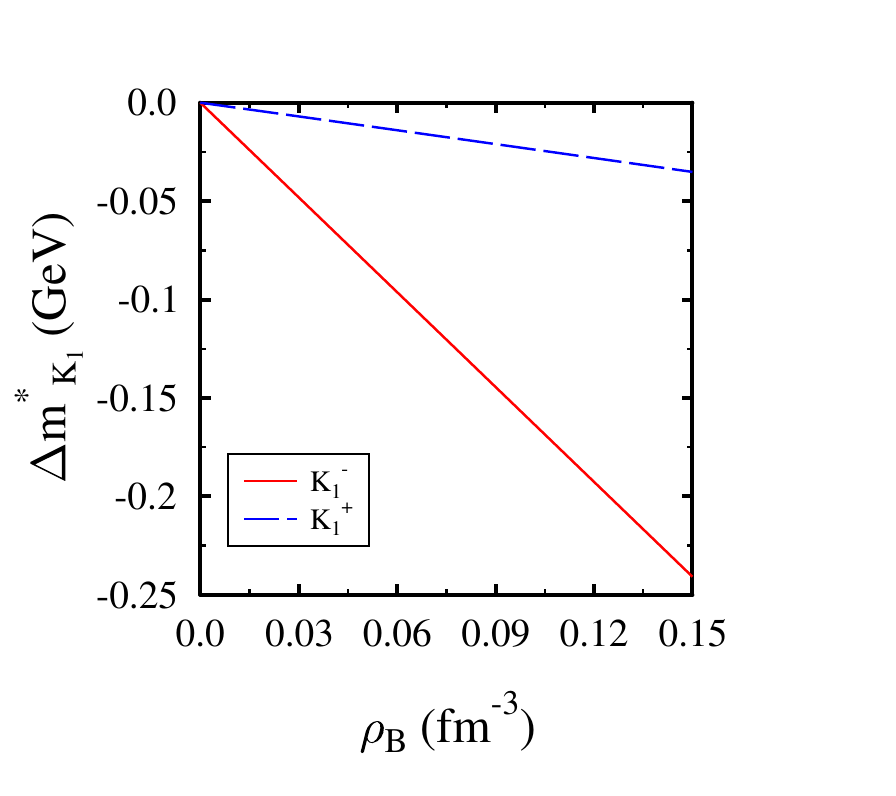}
		\caption{In-medium mass shift of $K_1^+$ and $K_1^-$.}
		\label{fig:gle}
	\end{figure}
	
	In fig. \ref{fig:gle}, we illustrate the in-medium mass of $K^+_1$ and $K_1^-$ mesons with nuclear density at zero temperature.  We observe a huge mass splitting between  charged partners of $K_1$ meson in symmetric nuclear medium, which increases with the rise in nuclear density. The attractive mass shift of $K_1^-$ meson grow significantly in the  medium whilst the mass shift of $K_1^+$ modifies feebly. Under the effect of finite temperature ($T$=0.150 GeV), we observed the change in mass to be very less for both charged partners. The results obtained here are in good agreement with the observations of QCD sum rules \cite{Song2019}. For better comparison,  the observations from the chiral hadronic model model + QCD sum rules (present investigation) and  QCD sum rules alone are compared in table \ref{table2}. In the solo QCD sum rules approach \cite{Song2019}, the quark and gluon condensates are computed under the linear density approximation. The  mass shift of $K_1^-$ can be experimentally anticipated from the hadronic decays $K_1 \rightarrow K \rho$ and $K_1 \rightarrow K^* \pi$ as well as  excitation function. The in-medium $K_1 N$ properties can also be studied  through the $K^-_1$ interactions with several nuclei at J-PARC \cite{Daum1981,Song2019}.
	

	\begin{table}
		\centering
		\begin{tabular}{ccccccc}
			\hline
			& $\Delta O^*_{K_1^-}$  & $\Delta m^*_{K_1^-}$ & $\Delta  {s^*_0{^-}}$& $\Delta O^*_{K_1^+}$ & $\Delta m^*_{K_1^+}$ & $\Delta  {s^*_0{^+}}$ \\
			\hline
			This work& -2.88$\times$10$^{-2}$  & -0.256 &-1.28  &-2.8$\times$10$^{-3}$&-0.0374&-0.247 \\
			
			QCD sum rules \cite{Song2019}& -3.09$\times$10$^{-2}$  & -0.249 &-1.25  &-2.72$\times$10$^{-3}$&-0.0348&-0.234 \\
			\hline
		\end{tabular}
		\caption{In-medium values  of shift in overlapping strength $\Delta O^*_{K_1^\pm}$ (GeV$^2$), mass $\Delta m^*_{K_1^\pm}$ (GeV), and, threshold parameter $\Delta  {s^*_0}^\pm$ (GeV$^2$) at $\rho_N$=0.16 fm$^{-3}$, $T$=0 MeV.}
		\label{table2}
	\end{table}

	\section{Conclusion}
	\label{sec:con}
	In the present investigation, using the  union of the  QCD sum rules and chiral hadronic model, we evaluated the medium modified  $K_1$ mass in the symmetric nuclear medium. We find that the charge symmetry does not hold in the nuclear medium for $K_1$ meson as the  medium influenced mass of $K_1^- (\bar u s)$ meson modifies significantly at nuclear saturation density whereas in-medium mass $K_1^+ (u \bar s)$ meson modifies less. 
	
	\section*{Acknowledgments}
	One of the authors, (R.K)  sincerely acknowledges the support towards this work from Ministry of Science and Human Resources Development (MHRD), Government of India via Institute fellowship under National Institute of Technology Jalandhar.

	\bibliography{kmeson_bib.bib}

	
\end{document}